\documentclass[useAMS,usenatbib]{mn2e}
\usepackage{times}
\title[]{Hydrodynamical wind on a magnetized ADAF with thermal conduction}
\author[ S. Abbassi, J. Ghanbari , and M. Ghasemnezhad]{S. Abbassi$^{1,2}$\thanks{E-mail:abbassi@mail.ipm.ir}, J. Ghanbari$^{3,4}$\thanks{E-mail: ghanbari@ferdowsi.um.ac.ir}
, and M. Ghasemnezhad$^3$ \\
$^{1}$School of Physics, Damghan University,
 P.O.Box 36715-364, Damghan, Iran\\
$^{2}$School of Astronomy, Institute for Research in Fundamental Sciences, P.O.Box 19395-5531, Tehran, Iran\\
$^{3}$Department of Physics, School of Sciences, Ferdowsi University of Mashhad, P.O.Box 91775-1436, Mashhad, Iran\\
$^{4}$Khayam Institute of Higher Education, P.O.Box 918974-7178, Mashhad, Iran}

\begin{document}
\date{}

\pagerange{\pageref{firstpage}--\pageref{lastpage}} \pubyear{2010}

\maketitle \label{firstpage}

\begin{abstract}
We examine the effects of a hydrodynamical wind on advection dominated accretion
flows with thermal conduction in the presence of a toroidal magnetic field under a self-similar treatment. 
The disk gas is assumed to be isothermal. For a steady state structure of such accretion flows a set of self similar 
solutions are presented. The mass-accretion rate $\dot{M}$ decreases with radius $r$ as 
$\dot{M}\propto r^{s+\frac{1}{2}}$, where $s$ is an arbitrary constant. We show that existence 
of wind will lead to enhance the accretion velocity. Cooling effects of outflows or 
winds are noticeable and should be taken into account for calculating luminosity and effective temperature of 
optically thin and thick ADAFs. Increasing the effect of wind, decreases the disk's temperature, 
because of energy flux which is taken away by winds. We will see that for a given set of input parameters, 
the solution reaches to a non-rotating limit at a specific value of $\phi_s$. The values of this limitation 
on $\phi_s$ will increase by adding $s$, wind parameter. In fact, the higher values of $s$ will allow the disk 
to have physical rotating solution for larger $\phi_s$.
\end{abstract}


\section{INTRODUCTION}

Accretions onto compact objects are the most energetic process in the universe. It is
believed that many astrophysical objects are powered by mass accretion on to black holes.
The standard geometrically thin, optically thick accretion disk model can successfully explain most
of the observational features in active galactic nuclei (AGN) and X-ray binaries (Shakura \& Sanyav 1973).

Accretion disks are an important ingredient in our current
understanding of many astrophysical systems on all
scales. Examples of presumed disks accretors include young stars,
compact objects in close binary system, active galactic
nuclei (AGN) and quasars (QSOs). There is also evidence that the
process of mass accretion via a disk is often and perhaps always
associated with mass loss from the disk in the form of a wind or
a jet. Mass loss appears to be a common phenomenon among
astrophysical accretion disk systems. These mass-loss mechanisms
are observed in microquasars, Young stellar objet and even from
brown dwarfs ( Ferrari 1998,Bally,Reipurth \& Davis 2007; Whelan et
al 2005). An outflow emanating from an accretion disk can act as a
sink for mass, angular momentum and energy and can therefore alter
the dissipation rates and effective temperature across the
disk(Knigge 1999). The accretion flows lose their mass by the
winds as they flow on to the central object. As a result of mass
loss,the accretion rate, $\dot{M}$, is no longer constant in
radius $r$. It is often expressed as $\dot{M}\propto r^{s}$ with
s being a constant of order unity (Blandford \& Begelman 1999).

Astrophysical jets and outflows emanating from accretion disks
have been extensively investigated by many researcher. Various
driving sources are proposed, including thermal, radiative and magnetic
ones. Traditionally the name of wind depends on its driving force.
In this study we will follow the hydrodynamical (thermal) wind which has been
discussed by many authors (e.g. Meier 1979, 1982, Fukue 1989, Takahara et. al 1989).

Accretion disk models have been extensively studied during the
past three decades (see Kato et al 1998 for a review). Besides introducing the
traditional standard disk by shakura and sunayev (1973), there are
new-type disks, such as advection-dominated accretion flows (ADAF)
or radiative inefficient accretion flows (RIAF) for a very small
mass-accretion disks (Narayan \& Yi 1994) and supercritical
accretion disk or so-called slim disks, for a very large
mass-accretion rate (Abramowicz et al 1998). ADAFs with winds or
outflows have been studied extensively during recent years, but
thermal conduction has been neglected in all these ADAFs
solutions with winds and toroidal magnetic field. Since the advection-dominated disks have high temperature,
the internal energy per particle is high. This is one of the reasons why advective cooling
overcomes radiative cooling. For the same reason, turbulent heat transport by conduction
in radial direction is non-negligible in the heat balance equation. Shadmehri (2008), Tanaka \& Menou (2006)
have studied the effect of hot accretion flow with thermal conduction. Shadmehri (2008)
has shown that thermal conduction opposes the rotational velocity,
but increases the temperature. In advected dominated inflow-outflow
solutions (ADIOS) it is assumed that the mass flow rate has a
power low dependence on radius with the power low index
$s$, treated as a parameter ( e.g. Blandford \& Begelman 1999).
Kitabatake , Fukue \& Matsumato (2002) studied supercritical
accretion disk with winds, though angular momentum loss of the
disks, because of the winds, has been neglected. In this paper, we
present self-similar solutions for ADAFs with thermal conduction, wind
and a toroidal magnetic field, in the basis of solutions presented by Akiziki \& Fukue (2006).
In \$2, we present assumptions and basic equations. Self-similar solutions are investigated in section 3. The aim of this work is
to consider possibility that winds in the presence of thermal
conduction and a toroidal magnetic field which has been largely
neglected before, could affect the global properties of hot
accretion flows substantially. In \$4 we show the results and give
discussions of the results.

\section{The Basic Equations}
We investigate the effect of mass outflow by the wind and mass accretion rate by viscosity
simultaneously. We consider an accretion disk that is stationary and axi-symmetric
 ($\frac{\partial}{\partial \phi}=\frac{\partial}{\partial
t}=0$) and geometrically thin $ \frac{H}{R}<1 $. In cylindrical
coordinates $(r,\varphi,z)$, we vertically integrate the flow
equations, also we suppose that all flow variables are only a
function of $r$. We ignore the relativistic effect and we use
Newtonian gravity. We adopt $\alpha$-prescription for viscosity of
rotating gas. For magnetic field we considered a toroidal
configuration.

The equation of continuity gives

\begin{equation}
\frac{\partial}{\partial r}(r\Sigma
v_{r})+\frac{1}{2\pi}\frac{\partial\dot{M_{w}}}{\partial r}=0
\end{equation}
where $v_{r}$ is the accretion velocity ($v_{r}<0$) and
$\Sigma=2\rho H$ is the surface density at a cylindrical radius
$r$. The mass loss rate by outflow/wind is represented by
$\dot{M_{w}}$. So
$$
 \dot{M_{w}}(r)=\int 4\pi r'\dot{m_{w}}(r')d r'
$$
 where $\dot{m_{w}}(r)$ is mass loss rate per unit area from each
disk face. On the other hand, we can rewrite the continuity equation:

\begin{equation}
\frac{1}{r}\frac{\partial}{\partial r}(r\Sigma v_r )=2 \dot{\rho}H
\end{equation}
where $\dot{\rho}$ the mass loss rate per unit volume and H is the
disk half-thickness.

The equation of motion in the radial direction is:
\begin{equation}
v_r\frac{\partial v_r}{\partial r}=\frac{v_\varphi^{2}}{r}-\frac{G
M_{\ast}}{r^{2}}-\frac{1}{\Sigma}\frac{d}{dr}(\Sigma
C_s^{2})-\frac{C_A^{2}}{r}-\frac{1}{2\Sigma}\frac{d}{dr}(\Sigma
C_A^{2})
\end{equation}
where $v_{\varphi}, C_s$ and $C_A$ are the rotational velocity of gas disk, sound and
Alfven velocities of the fluid respectively. Sound speed is defined as $C_s^2=\frac{p_{gas}}{\rho}$, $p_{gas}$
being the gas pressure and Alfven velocity is defined as $C_A^2=\frac{B_{\varphi}^2}{4\pi\rho}=\frac{2p_{mag}}{\rho}$,
where $p_{mag}$ being the magnetic pressure.

The integrated angular momentum equation over $z$ gives: (e.g., Shadmehri 2008)
\begin{equation}
r\Sigma v_r
\frac{d}{dr}(rv_\varphi)=\frac{d}{dr}(r^{3}\nu\Sigma\frac{d\Omega}{dr})-\frac{\Omega(lr)^{2}}{2\pi}\frac{d\dot{M_w}}{dr}
\end{equation}
where the last term on the right side represents angular
momentum carried by the outflowing material. Here, $l=0$
corresponds to a non-rotating wind and $l=1$ to outflowing
material that carries away the specific angular momentum (Knigge
1999). Also $\nu$ is kinematic viscosity coefficient and we assume:
\begin{equation}
\nu=\alpha C_s H
\end{equation}
where $\alpha$ is a constant less than unity. By integrating over
$z$ of the hydrostatic balance ,we have:
\begin{equation}
\frac{GM}{r^{3}}H^{2}=C_s^{2}[1+\frac{1}{2}(\frac{C_A}{C_s})^{2}]=(1+\beta)C_s^{2}
\end{equation}
where $\beta=\frac{P_{mag}}{P_{gas}}=\frac{1}{2}(\frac{C_A}{C_s})^{2}$
which shows the important of magnetic field pressure compared
to gas pressure. We will show the dynamical properties of the
disk for different values of $\beta$. Now we can write the energy
equation considering cooling and heating processes in an ADAF. We
assume the generated energy due to viscous dissipation and the
heat conduction into the volume are balanced by the
advection cooling and energy loss of outflow. Thus,
\begin{displaymath}
\frac{\Sigma v_r}{\gamma-1}\frac{dC_s^{2}}{dr}-2H v_r
C_s^{2}\frac{d\rho}{dr}=\frac{f\alpha\Sigma
C_s^{2}}{\Omega_k}r^{2}(\frac{d\Omega}{dr})^{2}-\frac{2H}{r}\frac{d}{dr}(rF_s)-
\end{displaymath}
\begin{equation}
\frac{1}{2}\eta\dot{m}_w(r)v_k^{2}(r)
\end{equation}
where the second term on right hand side represents energy
transfer due to the thermal conduction and $F_s=5\Phi_s \rho
C_s^{3}$ is the saturated conduction flux (Cowie \& Makee
1977). Dimensionless coefficient $\Phi_s$ is less than unity. Also,
the last term on right hand side of energy equation is the energy
loss due to wind or outflow (Knigge 1999). Depending on the energy
loss mechanisms, dimensionless parameter $\eta$ may change. In our case
we consider it as a free parameter in our models so that larger
$\eta$ corresponds to more energy extraction from the disk because of
outflows (Knigge 1999). Finally since we consider the toroidal
component for the global magnetic field of central stars, the
induction equation with field scape can be written as:
\begin{equation}
\frac{d}{dr}(V_r B_\varphi)=\dot{B_\varphi}
\end{equation}
where $\dot{B_\varphi}$ is the field scaping/creating rate due to
magnetic instability or dynamo effect.

\section{Self-Similar Solutions}
Self-similar solutions can not be able to describe the global behavior of the solutions,
because in this method there are not any boundary conditions which have been taken into account.
However as long as we are not interested in the solutions near the boundaries, such solution describe
correctly, true and useful asymptotically  behavior of the flow in the intermediate areas.

We assume that the physical properties are self-similar in the radial direction. In the
self-similar model the velocities are assumed to be expressed as follows:
\begin{equation}
v_r(r)=-C_1 \alpha v_k(r)
\end{equation}
\begin{equation}
v_\varphi(r)=C_2 v_k(r)
\end{equation}
\begin{equation}
C_s^{2}(r)=C_3 v_k^{2}(r)
\end{equation}
\begin{equation}
C_A^{2}(r)=\frac{B_\varphi^{2}(r)}{4\pi\rho(r)}=2\beta C_3v_k^{2}(r)
\end{equation}
where
\begin{equation}
v_k(r)=\sqrt{\frac{GM}{r}}
\end{equation}
and constants $C_1$,$C_2$ and $C_3$ will be determined later. From the
hydrostatic equation, we obtain the disk half-thickness H as:
\begin{equation}
\frac{H}{r}=\sqrt{C_3 (1+\beta)}=\tan\sigma
\end{equation}
Hence, a hot accretion flow with winds also has a conical
surface, whose opening (half-thickness) angle is $\sigma$.

We assume the surface density $\Sigma$ is in the form of:
\begin{equation}
\Sigma=\Sigma_0 r^{s}
\end{equation}

Note that the value of $s$ should be determined iteratively for consistency.
Then we assume that the power law index of the density $\rho$ in the radial direction
is constant regardless of $z$. Hence we set $\rho\propto r^{s-1}$

\begin{equation}
\dot{\rho}=\dot{\rho_0} r^{s-\frac{5}{2}}
\end{equation}
\begin{equation}
\dot{B_\varphi}=\dot{B_0} r^{\frac{s-5}{2}}
\end{equation}
\begin{equation}
\dot{M_w}=\dot{M_0} r^{s+\frac{1}{2}}
\end{equation}
\begin{equation}
\dot{m_w}=\dot{m_0} r^{s-\frac{3}{2}}
\end{equation}

It should be noted that, for a self-similar disk without any wind mass loss,
the suffix $s$ is $s=-\frac{1}{2}$.

By substituting the above self-similar solutions into the
dynamical equations of the system, we obtain the following dimensionless equations, to be solved for $C_1$, $C_2$ and
$C_3$:
\begin{equation}
\dot{\rho_0}=-(s+\frac{1}{2})\frac{C_1
\alpha\Sigma_0)}{2}\sqrt{\frac{GM_\ast}{(1+\beta)C_3}}
\end{equation}
\begin{equation}
H=\sqrt{(1+\beta)C_3}r
\end{equation}
\begin{equation}
-\frac{1}{2}C_1^{2}\alpha^{2}=C_2^{2}-1-[s-1+\beta(s+1)]C_3
\end{equation}
\begin{equation}
C_1=3(s+1)C_3+(s+\frac{1}{2})l^{2}\dot{m}
\end{equation}
\begin{equation}
(\frac{1}{\gamma-1}-\frac{1}{2})C_1 C_3 =\frac{9}{4}f C_3
C_2^{2}-\frac{5
\Phi_s}{\alpha}(s-\frac{1}{2})C_3^{\frac{3}{2}}-\frac{1}{8}\eta
\dot{m}
\end{equation}
\begin{equation}
\dot{m}=2C_1
\end{equation}
where $\dot{m}=\frac{\dot{M}_0w}{\pi\alpha\Sigma_0\sqrt{GM}}$ is
the nondimensional mass accretion rate. After algebraic
manipulations, we obtain a fourth order algebraic equation for
$C_1$:
\begin{displaymath}
D^{2}C_1^{4}+2DBC_1^{3}+(B^{2}+2D(E-1))C_1^{2}+
\end{displaymath}
\begin{equation}
(2B(E-1)-A^{2})C_1+(E-1)^{2}=0
\end{equation}
Where
\begin{equation}
D=\frac{1}{2}\alpha^{2}
\end{equation}
\begin{equation}
B={\frac{4}{9f}(\frac{1}{\gamma-1}-\frac{1}{2})-[s-1+\beta(s+1)][\frac{1-2(s+\frac{1}{2})l^{2}}{3(s+1)}]}
\end{equation}
\begin{equation}
A=\frac{20\Phi_s}{9f\alpha}(s-\frac{1}{2})[\frac{1-2(s+\frac{1}{2})l^{2}}{3(s+1)}]^{\frac{1}{2}}
\end{equation}
\begin{equation}
E=\frac{\eta}{3}\frac{(s+1)}{(1-2(s+\frac{1}{2})l^{2})}
\end{equation}

Abbassi et al. (2008) have solved these equations when $s=-\frac{1}{2}$, because
they did not have any wind or mass-loss in their model. This
algebraic equation shows that the variable $C_1$ which determines
the behavior of radial velocity depends only on the
$\alpha$, $\Phi_s$, $\beta$, $f$, $s$ and $\eta$. Other flows quantities such as
$C_2$ and $C_3$ can be obtained easily from $C_1$:
\begin{displaymath}
C_2^{2}=\frac{4 C_1}{9f}[\frac{1}{\gamma-1}-\frac{1}{2}]
\end{displaymath}
\begin{equation}
+\frac{20\Phi_s}{f\alpha}(s-\frac{1}{2})[\frac{1-2(s+\frac{1}{2})l^{2}}{3(s+1)}]^{\frac{1}{2}}
C_1^{\frac{1}{2}}+\frac{\eta}{3}(\frac{s+1}{1-2(s+\frac{1}{2})l^{2}})
\end{equation}
\begin{equation}
C_3=C_1(\frac{1-2(s+\frac{1}{2})l^{2}}{3(s+1)})
\end{equation}

We can solve these simple equations numerically and clearly just physical solutions
can be interpreted. Without wind, thermal conduction and magnetic field, $s=l=\eta=\phi_s=\beta=0$,
the equations and their similarity solutions reduce to the standard ADAFs solution ( Narayan \& Yi 1994).
Also without wind they reduce to Abbassi et al. (2008). Now we can analysis the behavior of solutions.

\section{Results}

\input{epsf}
\epsfxsize=3in \epsfysize=4.1in
\begin{figure}
\centerline{\epsffile{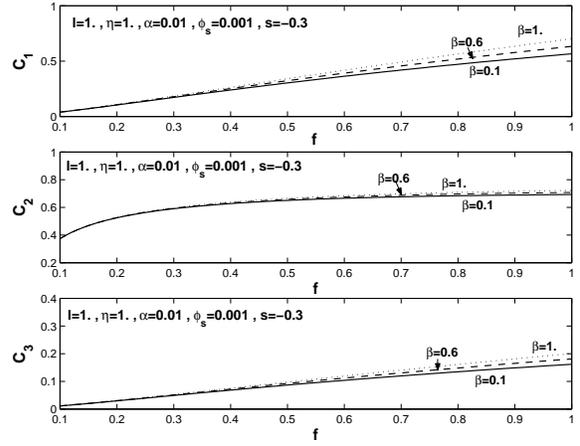}}
 \caption{Numerical coefficient $C_i$s as a function of advection parameter $f$ for several values of
 $\beta$, magnetic field strength. All of this figures was set up for $s=-0.3$, $\alpha=0.01$, $\phi=0.001$ and
 $\l=\eta=1$.   }

 \label{figure1}
\end{figure}
\input{epsf}
\epsfxsize=3in \epsfysize=4.1in
\begin{figure}
\centerline{\epsffile{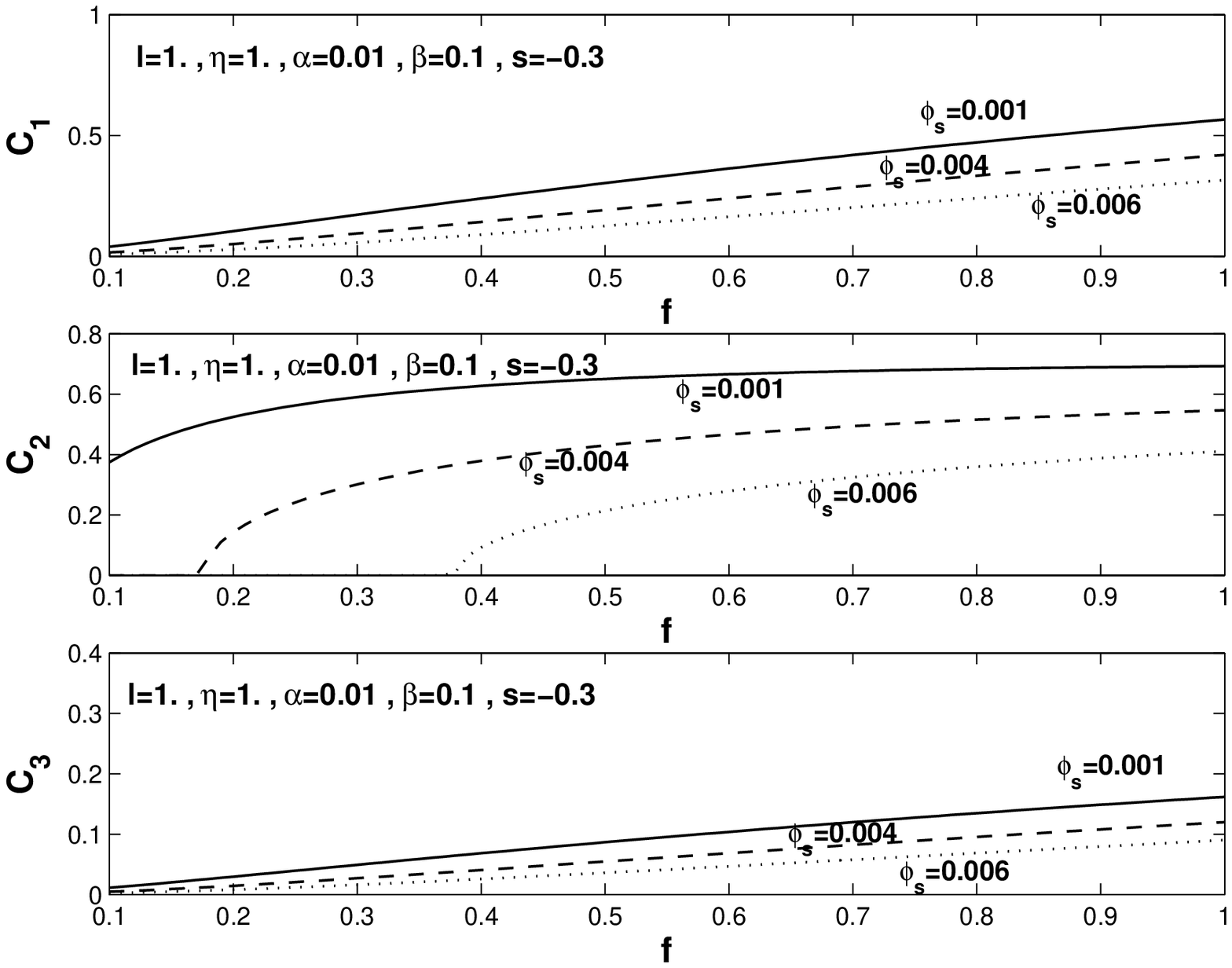}}
 \caption{ Numerical coefficient $C_i$s as a function of advection parameter $f$ for several values of
 $\phi_s$, thermal conduction parameter. All of this figures was set up for $s=-0.3$, $\alpha=0.01$, $\beta=0.1$ and
 $\l=\eta=1$.  }
 \label{figure2}
\end{figure}
\input{epsf}
\epsfxsize=3in \epsfysize=4.1in
\begin{figure}
\centerline{\epsffile{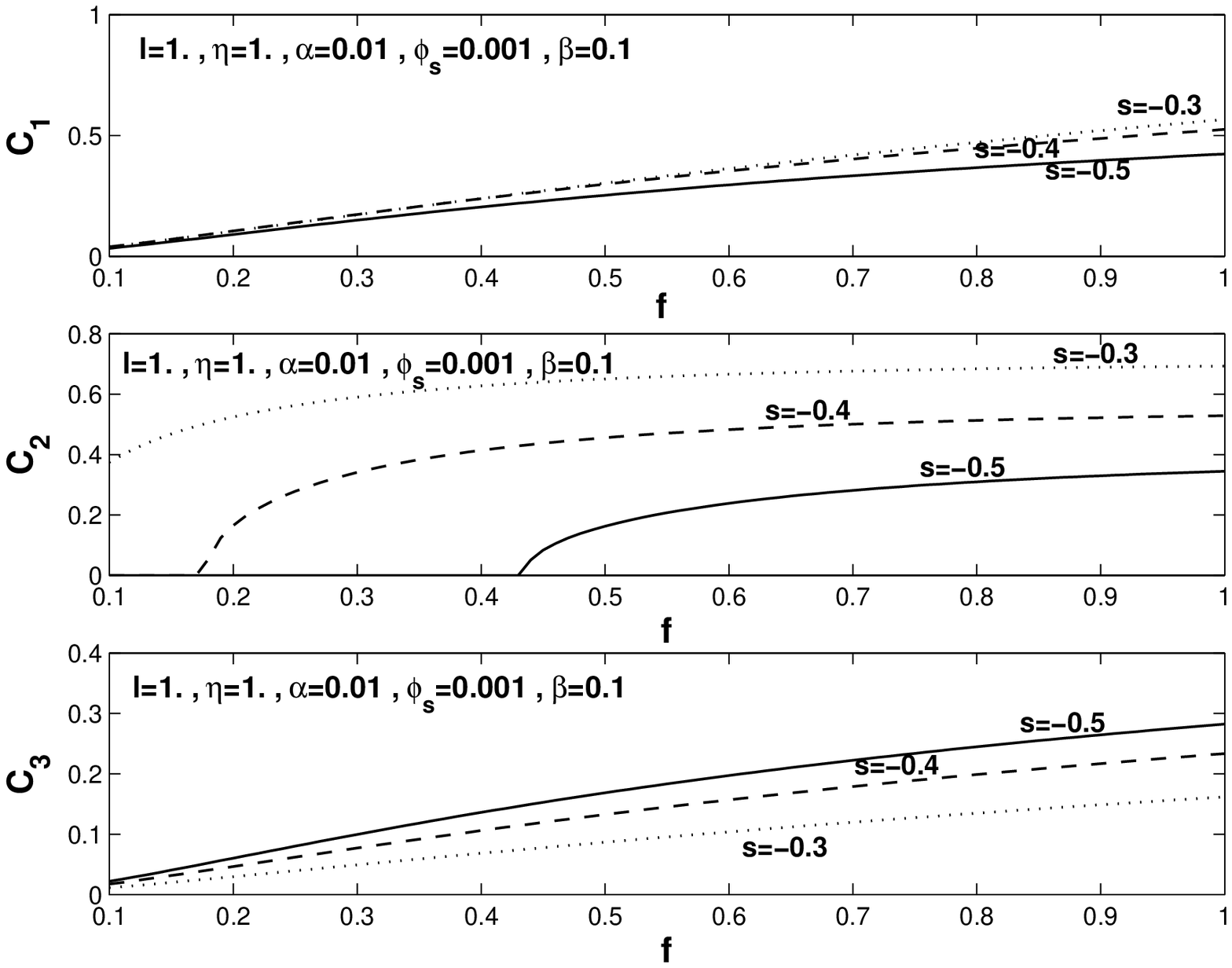}}
 \caption{Numerical coefficient $C_i$s as a function of advection parameter $f$ for several values of
 $s$, wind parameter. All of this figures was set up for $\phi_s=0.001$, $\alpha=0.01$, $\beta=0.1$ and
 $\l=\eta=1$.  }
 \label{figure3}
\end{figure}

\input{epsf}
\epsfxsize=3in \epsfysize=4.1in
\begin{figure}
\centerline{\epsffile{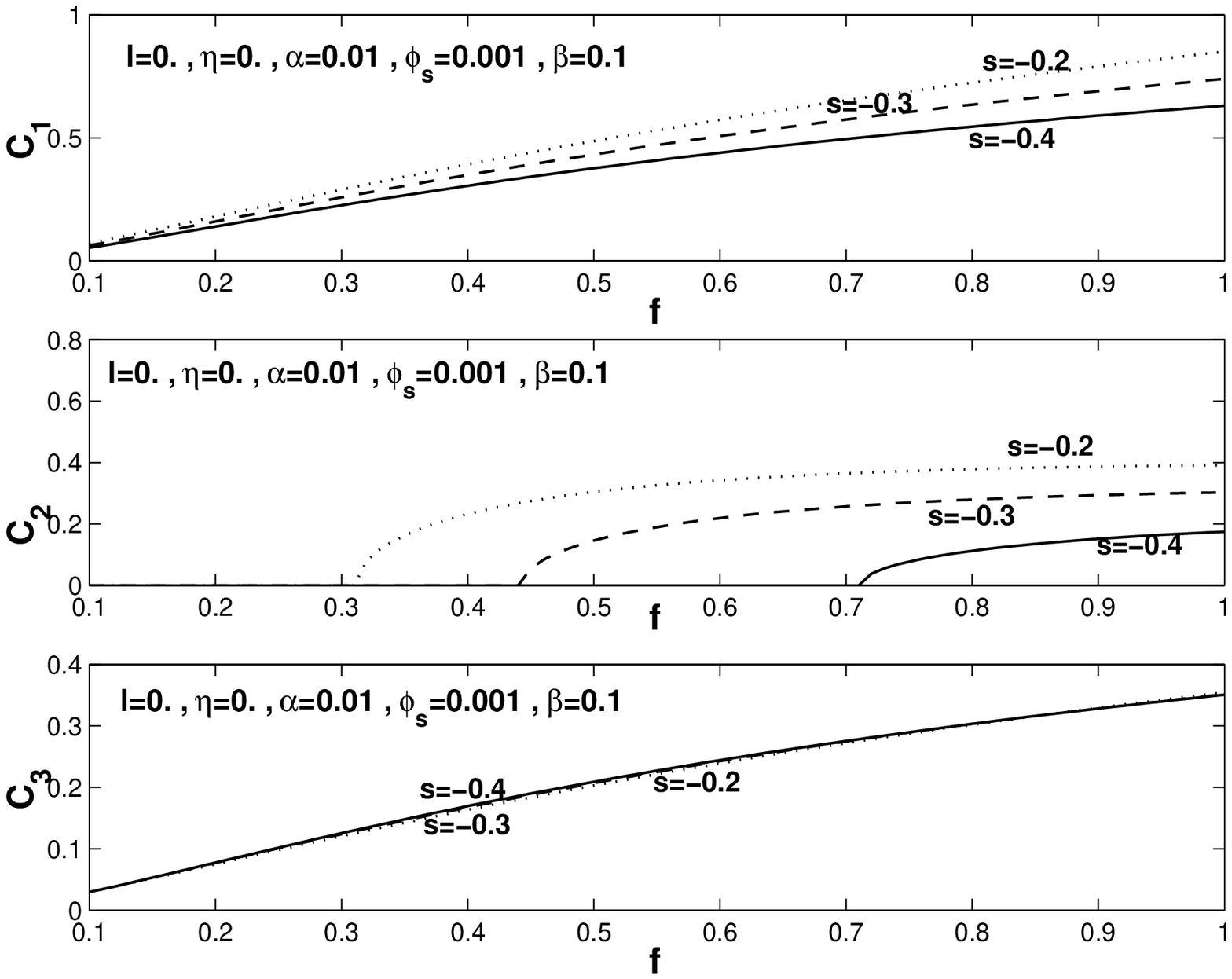}}
 \caption{Numerical coefficient $C_i$s as a function of advection parameter $f$ for several values of
 $s$, wind parameter. All of this figures was set up for $\phi_s=0.001$, $\alpha=0.01$, $\beta=0.1$ and
 $\l=0$, $\eta=0$. }
 \label{figure4}
\end{figure}

\input{epsf}
\epsfxsize=3in \epsfysize=4.1in
\begin{figure}
\centerline{\epsffile{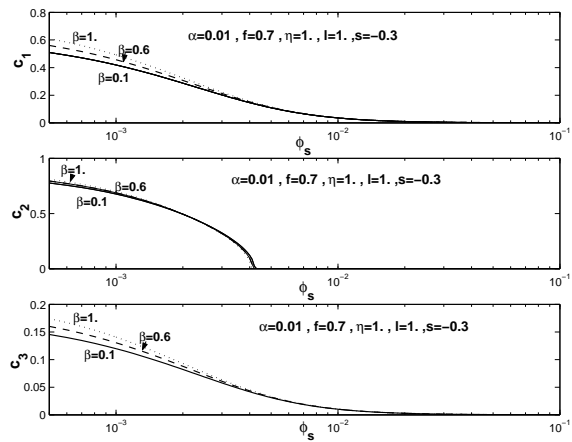}}
 \caption{Numerical coefficient $C_i$s as a function of thermal conduction parameter $\phi_s$ for several values of
 $\beta$, magnetic field parameter. All of this figures was set up for $f=0.7$, $\alpha=0.001$, $s=-0.3$ and
 $\l=1$, $\eta=1$. }
 \label{figure4}
\end{figure}

\input{epsf}
\epsfxsize=3in \epsfysize=4.1in
\begin{figure}
\centerline{\epsffile{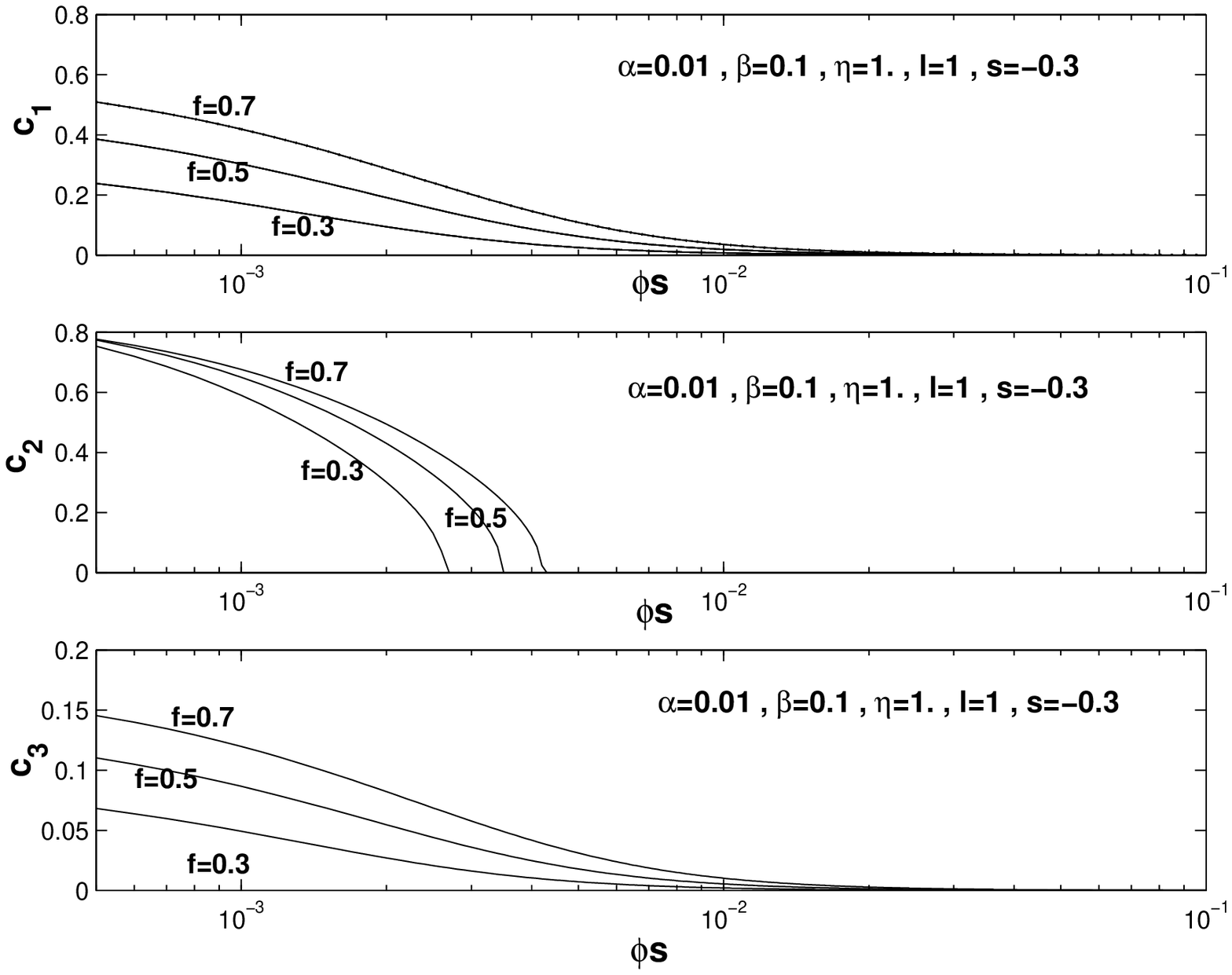}}
 \caption{Numerical coefficient $C_i$s as a function of thermal conduction parameter $\phi_s$ for several values of
 $f$, advection parameter. All of this figures was set up for $\alpha=0.01$, $\beta=0.1$, $s=-0.s$ and
 $\l=1$, $\eta=1$. }
 \label{figure4}
\end{figure}

\input{epsf}
\epsfxsize=3in \epsfysize=4.1in
\begin{figure}
\centerline{\epsffile{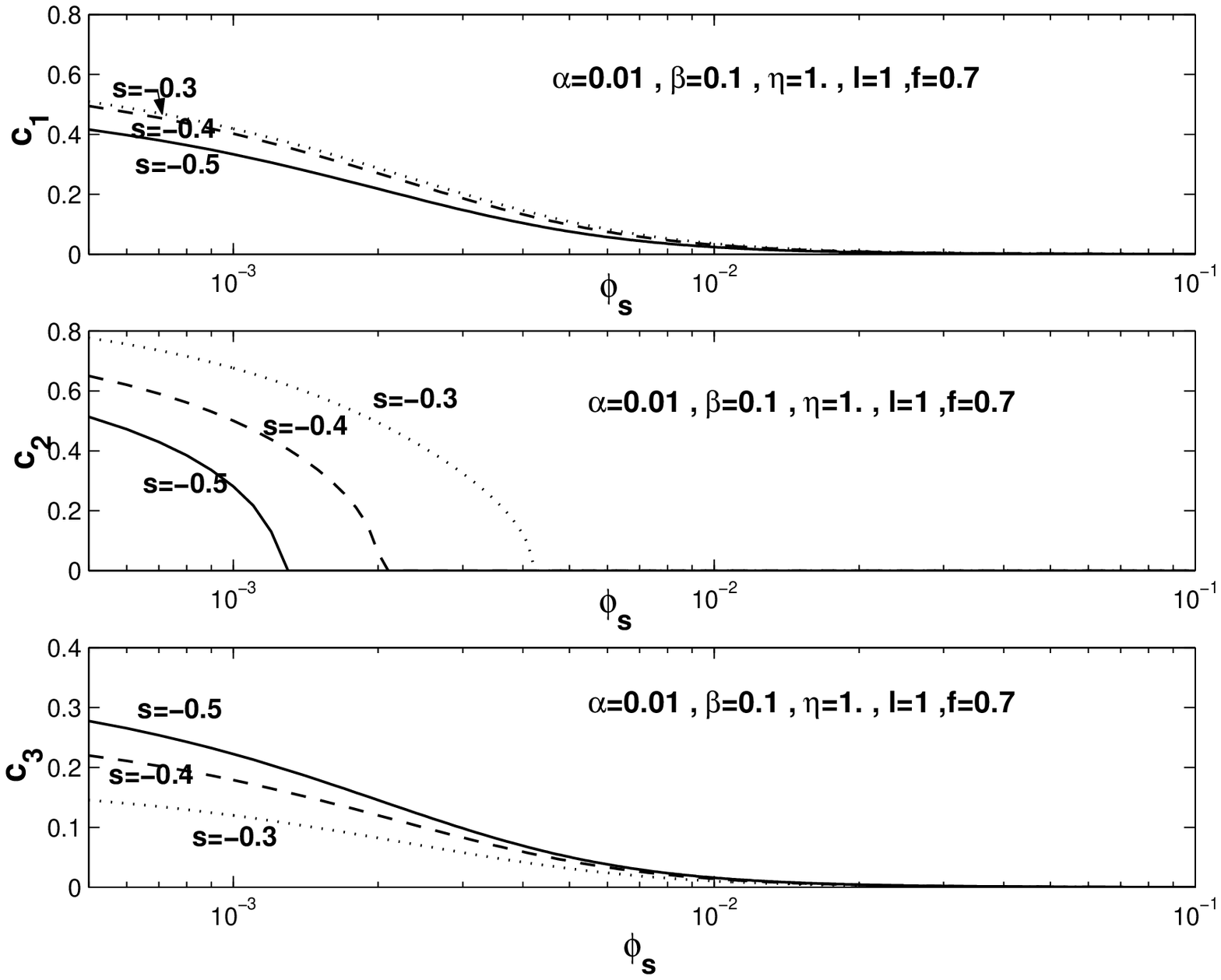}}
 \caption{Numerical coefficient $C_i$s as a function of thermal conduction parameter $\phi_s$ for several values of
 $s$, wind parameter. All of this figures was set up for $\alpha=0.01$, $\beta=0.1$, $f=0.7$ and
 $\l=1$, $\eta=1$. }
 \label{figure4}
\end{figure}

Now we can analysis the behavior of solutions in the presence of outflow,
thermal conduction and magnetic field. Our primary goal is to investigate these effects
via parameters $s, \l, \eta, \phi_s$ and $\beta$.

This Algebraic equation shows that the variable $C_1$ which
determines the behavior of radial velocity depends only on the
$\alpha$, $\phi$, $\beta$, $s$ and $f$. Other flows quantities such as
$C_2$ and $C_3$ can be obtained easily from $C_1$ via the above equations.
 The parameters of the model are the ratio of specific heats $\gamma$, the standard viscose parameter
$\alpha$, the energy-advection parameter $f$, degree of
magnetic pressure to gas pressure $\beta$ and $s$ which determines the outflow from the disks.

Figure 1. shows the coefficients $C_i$s in terms of advection parameter $f$ for different values of $\beta$.
In the upper panel we have shown $C_1$ as a function of advection parameter. It represents the behavior of radial flows
of accretion materials. Although in ADAFs the radial velocity is generally sub-Keplerian, it becomes larger by increasing $f$.
By adding $\beta$, which indicates the role of magnetic field on the dynamics of accretion disks, we will see that
the radial flow will increases. On the other hand the radial flow increases when the toroidal magnetic field becomes large which
agrees with Abbassi et al. (2008)'s results. This is because, the magnetic tension term dominates the magnetic pressure term in
the radial momentum equation, which assists the radial velocity of accretion flows. The middle panel, shows the coefficient $C_2$
which is the ratio of the rotation velocity to the Keplerian velocity as a function of advection parameter for different values
of $\beta$. We can see that by adding the magnetic field influences (adding $\beta$), the rotation velocity will decreases.
This is because of that the disk should rotate faster than the case without the magnetic field which results the magnetic tension.
In the lower panel, we have plotted $C_3$, the coefficient of squared sound speed as a function of $f$.
As the advection parameters becomes large, the sound speed and therefore the disk's thickness becomes large. Sound speed also depends
on the magnetic field influences. By adding magnetic field influence we will see that sound speed increases as well as vertical thickness
of accretion flows.

In Fig. 2 we investigate the role of saturated thermal conduction in radial, toroidal and sound speed of accretion flows.
Increasing the thermal conduction coefficient $\phi_s$, will decrease the radial velocity. It will have a large effect on the
rotational velocity of accretion flows (Middle panel). It will decrease the sound speed and therefore vertical thickness of the disks
too (lower panel).

To show the behavior of the solution respect to the wind influences we have plotted disk's physical quantities for different
values of $s$ and $\l$ in Fig.3 and Fig.4. The value of $s$ measures the strength of outflows, and larger values
of $s$ denotes strong outflows. Each curve is labeled with its corresponding $s$. We can see that
ADAFs with wind rotates more quickly than those without wind, radial flows of the accretion materials also increases for larger $s$.
Strong-wind models have a lower vertical thickness compare to the weak-wind model because by adding $s$ we can see that $C_3$
decreases which means that the sound speed and vertical thickness decreases as well.

There are some limitations for having hot disks with sufficient low $f$. Middle panel in Fig.3 shows that there is no physical solution for
ADAFs with $f<0.4$ for the case of no-wind. But by adding $s$ we can see that this limitation can be canceled. The outflow and wind play as a cooling
agent while thermal conduction provides an extra heating source so these two factors affects the dynamical behavior of the flows.
While effects of wind and thermal conduction on the profiles of sound speed and vertical thickness of the disks is similar, their
effects on the profiles of radial and rotational velocities oppose each other. Both of them leads to decrease the sound speed and therefore
vertical thickness of the disks decreases. They lead to enhanced radial and rotational velocities.

In the lower panel of Fig.3 and Fig.4 we have plotted $C_3$ for a sequence of $f$. They show two cases, $\l=\eta=0$ and $\l=\eta=1$.
In the case of $\l=\eta=0$, variations of $s$ do not affect on $C_3$ profiles ( and vertical thickness of the disks); while in the case $\l=\eta=1$ it decreases when we increase the wind effect. Higher $\eta$ means more energy flux is taking away by winds and $\l$ will account for the
angular momentum carry out by the wind. So solutions with no-rotating wind $\l=0$ and no energetic wind $\eta=0$ was plotted in Fig.4.

In Fig.5, Fig.6 and Fig.7 we have plotted the coefficient $C_i$s with thermal conduction parameter $\phi_s$ for different values of $\beta$, $f$ and $s$. Tanaka \& Menou (2006) have shown that for a very small $\phi_s$ their solutions coincide the original ADAFs solution Narayan \& Yi (1994); But by adding saturated conduction parameter, $\phi_s$, the effect of thermal conduction can be better seen when we approach $\phi_s\sim 0.001-0.01$. So we have plotted our solutions in this range.  We can see that for a given set of input parameters, the solution reaches to a non-rotating limit at a specific value of $\phi_s$. We can not extend the solution for values of this limit for $\phi_s$. Because Eq.31 gives a negative value for $C_2^2$ which is clearly unphysical. The values of this limitation on $\phi_s$ will increase by adding $s$, wind parameter. In fact, the higher values of $s$ will allow the disk to have physical rotating solution for larger $\phi_s$, Fig.7 middle panel. In Fig.5 we can see that magnetic field has an important role on the radial velocity and the sound speed of the flow but it can not modify the rotation velocity which are compatible with Akizuki \& Fukue (2006), Fukue (2004) and Abbassi et al (2008). The sound speed and radial infall velocity both increase with magnitude of thermal conduction.

As the level of thermal conduction is increased, more and more heat flows from hot inner regions, resulting in a local increase of the gas
temperature relative to original ADAFs solutions.  Simultaneity, the gas adjust its angular velocity and increases its inflow speed to conserve
its momentum balance. We have found that the level of advection, $f$, will modified the dynamical structure of disk significantly. The breakdown
the solutions, when $C_s^2\rightarrow 0$, occurs at lower values of $f$. Clearly Thermal conduction can significantly affect the structure and the
properties of the hot accretion flows. To better understand the role of thermal conduction it is better to study it in a more realistic
two-dimensions investigations, without height integrations.

Our results that shows out flow (large $s$) in ADAFs will modify the dynamical quantities of accretion flows significantly.
Recently Kawabata \& Mineshige (2009) have shown that thermally driven wind with a power law formula is expected for such hot ADAFs disks.
They have also shown that the ratio of the outflow to the accretion rate is approximately unity for $\alpha\leq 0.1$. So our investigation should
have significant role to understand the physics of radiation inefficient accretion flows RIAFs or ADAFs.

However such kind of solutions enable us to investigate easily all possible cases for hot accretion flows with thermal conduction, wind, B-field and any
combinations of these physical mechanisms. This simple toy model will prepare a frame work to compare these effects.

\section{GENERAL PROPERTIES OF ACCRETION FLOWS}
In the pervious sections we have introduced solutions of advection type
accretion flow with winds in the presence of thermal conduction bathed in a toroidal
magnetic field. In this section we will discuss the general properties of accretion flow and we will investigate
possible effects of wind, thermal conduction and magnetic field on the
physical quantities of accretion flows.

Using the self-similar solution we will estimate the mass-accretion rate
as:
\begin{displaymath}
\dot{M}=-2\pi r\Sigma v_r= 2\pi\Sigma_0 c_1\alpha\sqrt{GM}r^{s+1/2}
\end{displaymath}
\begin{equation}
~~~= {\dot{M}_{out}}(\frac{r}{r_{out}})^{s+1/2}
\end{equation}
where $r_{out}$ is the disks outer radius and $\dot{M}_{out}$
is mass-accretion rate there. In the case of accretion disk with no wind,
$s=-1/2$ the accretion rate is independent of the radius, while for those
with wind, $s>-1/2$, the accretion rate decreases with radius as we expect. Because
winds start from various radii, the mass loss rate is not constant but depends on the radius.
As a result, some parts of accretion materials are not concentrated at the center, but are dilated
over a wide space. According to wind mass loss, the accretion rate decreases with radius as we expected.

With a simple calculation we will see that a significant amount of accretion materials flung away via the wind and mass loss.
Only $1-10\%$ of $M_{out}$ is ultimately accreted onto the central accretor. Using the above expression for accretion
rate, we have:

\begin{displaymath}
\frac{\dot{M}_{in}}{\dot{M}_{out}}\sim(\frac{r_{in}}{r_{out}})^{s+1/2}
\end{displaymath}
if we estimate $r_{in}\sim10^{-3}r_{out}$ we finally have for $s=1/2$:
\begin{displaymath}
\frac{\dot{M}_{in}}{\dot{M}_{out}}\sim 10^{-3}
\end{displaymath}

We can show the radial behavior of temperature of the present self-similar ADAFs disks with toroidal magnetic field and outflow. These advection-dominated accretion flows occurs in two regimes depending on their mass accretion rate and optical depth. Optical depth of accretion flows are highly depends on their accretion rate. In a high-mass accretion rate, the optical depth becomes very high and the radiation generated by accretion flows can be trapped within the disk. This type of accretion disks is known as \textit{optically-thick} or slim disks. In the limit of low-mass accretion rates, the disk becomes \textit{optically thin}. In this case, the cooling time of accretion flows is longer than the accretion time scale. The energy generated by accretion flows therefore remains mostly in the disks and the disks can not radiate their energy efficiently. We will examine the influence of the wind and thermal conduction on the radial appearances of the temperature in these two cases. In the optically thin case, where the gas pressure is dominated, we can adopt the ideal gas law to estimate the effective temperature as (Akizuki \& Fukue 2006):
\begin{equation}
\frac{R}{\bar{\mu}}T=c_S^2=\frac{GM}{r}
\end{equation}
where $T$ is the gas temperature, $R$ gas constant and $\bar{\mu}$ the mean molecular weight. If we use
($GM=r_g c$), where $r_g$ and $c$ are the Schwartzshild radius and light speed respectively, the temperature gradient is expected as:
\begin{equation}
T=c_3\frac{c^2/2}{R/\bar{\mu}}\frac{r_g}{r}
\end{equation}
which means that $T\propto\frac{c_3}{r}$. This has similar form ( radial dependency) with non magnetic case,
but coefficient $C_3$ implicitly depends on magnetic field $\beta$, outflow effect $s$, advection parameter $f$ and the effect of thermal conduction
$\phi$. For the case of strong wind $C_3$ decreases, Fig.3, so the surface temperature and vertical thickness of
 the disk will decrease. Thermal conduction has the same effect, Fig.2, while the magnetic field will increase the
 surface temperature and vertical thickness of the disk.

It should be emphasized that the radiative appearance of optically thin disks such as $L_{\nu}$, can not be calculated easily.
In this case it should demonstrate the importance of advective cooling in optically-thin disks. In order to clarify this problem we should show that cooling and heating how change with $r$, $\Sigma$ and $\dot{M}$. Radiative cooling generally has very complicated parameter dependencies. In the optically thin cases, emission from the gas is not a blackbody continuum. Bremsstrahlung cooling by non-relativistic or relativistic, Synchrotron and Compton cooling may have possible role to reproduce emission spectra.

In the optically thick case, where the radiation pressure dominated, sound speed is related to radiation pressure.
We can write the average flux $F$ as:
\begin{equation}
 F=\sigma T_c^4=\frac{3c}{8H}\Pi_{gas}=\frac{3}{8}c\Sigma_0\sqrt{\frac{c_3}{1+\beta}}GM r^{s-2}
\end{equation}
where $\Pi_{gas}=\Sigma c_s^2$ is the height-integrated gas pressure, $T_c$ disk central temperature and $\sigma$ the Stefan-Boltzman constant. In the moderately to the strongly magnetized $\beta\geq 1$ cases the magnetic pressure is comparable with gas pressure and it should take into account:
\begin{displaymath}
\Pi=\Pi_{gas}+\Pi_{mag}=(1+\beta)\Pi_{gas}
\end{displaymath}
so the calculation should be modified by multiply to $1+\beta$. For optically thick case the optical thickness of the disk in the vertical direction is:
\begin{displaymath}
\tau=\frac{1}{2}\kappa\Sigma=\frac{1}{2}\kappa\Sigma_0 r^{s},
\end{displaymath}
where $\kappa$ is the electron-scattering opacity. So we can calculate the effective flux and effective temperature of the disk surface as:
\begin{equation}
\sigma T_{eff}^4=\frac{\sigma T_c^4}{\tau}=\frac{3c}{4\kappa}\sqrt{\frac{c_3}{1+\beta}}\frac{GM}{r^2}=\frac{3}{4}\sqrt{\frac{c_3}{1+\beta}}
\frac{L_E}{4\pi r^2},
\end{equation}
\begin{equation}
 T_{eff}=(\frac{3L_E}{16\pi\sigma}\sqrt{\frac{c_3}{1+\beta}})^{1/4} r^{-1/2}
\end{equation}
where $L_E=4\pi c\frac{GM}{\kappa}$ is the Eddington Luminosity of the central object. If we integrate these equations radially
we have the disk luminosity as:
\begin{equation}
L_{disk}=\frac{3}{4}\sqrt{\frac{c_3}{1+\beta}}L_{E}\ln \frac{r_{out}}{r_{in}},
\end{equation}

As this equation shows, the optically thick disk's luminosity is affected by magnetic field explicitly, $\beta$,  but it would be affected by
outflow, thermal conduction and viscosity through the $C_3$ implicitly.

It should be emphasized that the Luminosity and effective temperature of the disk, $L_{disk}$ and $T_{eff}$, are not affected by
the mass loss through outflow (there are not any s dependencies). But wind would affect on the radiative appearance of the disk through the
$c_3$ in these formulas, implicitly. The average flux decreases all over the disk when we have mass loss outflow compared with the case of
no mass loss. The surface density and there fore optical depth decrease for the mass loss case. So we can see the deep inside of the disk.

\section{SUMMERY AND CONCLUSION }
In this paper we have studied accretion disks around a black hole in an advection-dominated regime in
the presence of a toroidal magnetic field. Thermal conduction and wind as a energy sources and angular momentum transport were adopted.
We have presented the results of self-similar solutions to show the effects of various physics behavior on the dynamical quantities of
accretion flows. We adopt the solution presented by Akizuki \& Fukue (2006), Fukue (2004) and Abbassi et al (2008) to present dynamical behavior of the advection dominated accretion flows. Some approximations were made in order to simplify the main equations. We assume an axially symmetric, static disk
with $\alpha$-prescription of viscosity. We ignored the relativistic effects and the self-gravity of the disks. Considering weakly collisional
nature of hot accretion flow ( Tanaka \& Menou 2006, Abbassi et al. 2008), a saturated form of thermal conduction was adopted as a possible
mechanism for energy transportation was adopted. We have accounted for this possibility by allowing the saturated thermal conduction
constant, $\phi_s$, to vary in our solutions.

Theoretical arguments and observations suggest the mass loss via wind in RIAFs. By assuming that the flow has
self-similarity structure in the radial direction, a power-law wind is  adopted in our model. Using some assumptions, we made a
simplified toy model which included the effect of thermal conduction, wind and B-field. In this toy model, we can easily investigate
its possible combine effects on the dynamical quantities of the fluid.

We have shown that strong wind could have lower temperature which are satisfied with the results were presented by Kawabata \& Mineshige 2009, which
could make significant differences in the observed flux compared to standard ADAFs. The most important finding of our simple self-similar solutions
is that the accreting flow is strongly affected not only by mass loss but also by energy loss by the wind.

There are some limitations in our solutions. One of them is that the magnetic field also has an important role to produce wind (magnetically driven wind). To having X-wind, pure toroidal B-field in not enough and the disk should have a $z$-component B-field. So our case is not a good model for magnetically driven disks. The other limitation of our solutions is the anisotropic character of conduction in the presence of magnetic field. Balbus (2001)
has argued that the dynamical structure the hot flows could be affected by anisotropic character of thermal conduction in the presence of B-field.

Although our preliminarily self-similar solutions are too simplified, but clearly improve our understanding of the  physics of advection-dominated accretion flows around a black hole. For having a realistic picture of accretion flow a global solution needed rather than the self-similar solution.


\begin{thebibliography}{}
\bibitem{} Abramowicz M. A., Czerny B., Lasota J. P., Szuszkiewicz
E., 1988, ApJ, 332, 646
\bibitem{} Abbassi S., Ghanbari J., Najjar S., 2008, MNRAS, 388, 663
\bibitem{} Abbassi S., Ghanbari J., Salehi  F., 2006, A\&A, 460, 357A
\bibitem{} Akizuki C. \& Fukue J., 2006, PASJ, 58, 1073
\bibitem{} Bally J., Reipurth B., Davis C. G., 2007, Proto-stars and planets, University Arizona press, p215-230
\bibitem{} Balbus S., 2001, ApJ, 562,909
\bibitem{} Blandford R. D., Begelman M, 1999, MNRAS, L1
\bibitem{} Bisnovatyi-Kogan G. S., Ruzmaikin A. A., 1976, Ap\&SS, 42, 401
\bibitem{} Cowie, L. L., Mckee, C.F., 1977, ApJ, 275, 641
\bibitem{} Ferrari A., 1998, ARA\&A, 36, 539
\bibitem{} Fukue J., 1989, PASJ, 41, 123
\bibitem{} Fukue J., 2004, PASJ, 56, 569
\bibitem{} Kitabatake E., Fukue J., Matsumoto K., 2002, PASJ, 54, 235
\bibitem{} Kato S., Fukue J., Mineshige S., 2008, Black Hole Accretion Disks, Kyoto University Press
\bibitem{} Kawabata R., Mineshige S., 2009, PASJ, 61, 1135
\bibitem{} Knigge C., 1999, MNRAS, 309, 409
\bibitem{} Meier D. L., 1979, ApJ, 233, 664
\bibitem{} Meier D. L., 1982, ApJ, 256, 706
\bibitem{} Narayan R., \& Yi, I. 1994, ApJ, 428, L13
\bibitem{} Narayan R., \& Yi, I. 1995a, ApJ, 444, 238
\bibitem{} Narayan R., \& Yi, I. 1995b, ApJ, 452, 710
\bibitem{} Narayan R., Maclintock J.E., Yi I., 1996, ApJ, 457, 821
\bibitem{} Shadmehri, M. 2008, AP\&SS, 317, 201S
\bibitem{} Shakura N. I., \& Sunyaev, R.A. 1973, A\&A,  24, 337
\bibitem{} Takahara F., Rosner R., Kusnose M., 1989, ApJ, 346, 122
\bibitem{} Tanaka T., Menou K.,2006, APJ, 649, 345
\bibitem{} Whelan E, et al, Nature, 439, 652



\end{thebibliography}
\end{document}